\documentclass[tinyml]{acmart}
\pdfoutput=1
\usepackage{adjustbox}

\AtBeginDocument{%
  \providecommand\BibTeX{{%
    \normalfont B\kern-0.5em{\scshape i\kern-0.25em b}\kern-0.8em\TeX}}}

\setcopyright{rightsretained}
\copyrightyear{2022}
\acmYear{2022}

\begin{document}

% \settopmatter{printacmref=true}
% \settopmatter{printfolios=false}
\title{Millimeter-Scale Ultra-Low-Power Imaging System for Intelligent Edge Monitoring}

\settopmatter{authorsperrow=4}
\author{Andrea Bejarano-Carbo}
\orcid{0000-0002-7269-4545}
\email{bejarano@umich.edu}
\affiliation{}

\author{Hyochan An}
\orcid{0000-0002-6322-025X}
\affiliation{}
\email{hyochan@umich.edu}

\author{Kyojin Choo}
\orcid{0000-0001-8119-094X}
\email{kjchoo@umich.edu}
\affiliation{}

\author{Shiyu Liu}
\email{shiyuliu@umich.edu}
\affiliation{}

\author{Qirui Zhang}
\email{qiruizh@umich.edu}
\orcid{0000-0001-8113-3558}
\affiliation{}

\author{Dennis Sylvester}
\orcid{0000-0003-2598-0458}
\affiliation{}
\email{dmcs@umich.edu}

\author{David Blaauw}
\orcid{0000-0001-6744-7075}
\email{blaauw@umich.edu}
\affiliation{}

\author{Hun-Seok Kim}
\orcid{0000-0002-6658-5502}
\affiliation{}
\email{hunseok@umich.edu}

\author{}
\affiliation{}
\author{}

\author{}
\affiliation{}
\author{}
\author{}
\affiliation{}
\author{}

\affiliation{
\institution{University of Michigan}
  \city{Ann Arbor}
  \state{Michigan}
  \country{USA}
}

\renewcommand{\shortauthors}{Bejarano-Carbo, et al.}

\begin{abstract}
Millimeter-scale embedded sensing systems have unique advantages over larger devices as they are able to capture, analyze, store, and transmit data at the source while being unobtrusive and covert. However, area-constrained systems pose several challenges, including a tight energy budget and peak power, limited data storage, costly wireless communication, and physical integration at a miniature scale. This paper proposes a novel 6.7$\times$7$\times$5mm imaging system with deep-learning and image processing capabilities for intelligent edge applications, and is demonstrated in a home-surveillance scenario. The system is implemented by vertically stacking custom ultra-low-power (ULP) ICs and uses techniques such as dynamic behavior-specific power management, hierarchical event detection, and a combination of data compression methods. It demonstrates a new image-correcting neural network that compensates for non-idealities caused by a mm-scale lens and ULP front-end. The system can store 74 frames or offload data wirelessly, consuming 49.6$\upmu$W on average for an expected battery lifetime of 7 days.

\vspace{-2mm}
\end{abstract}


\maketitle
\vspace{-2.5mm}
\section{Introduction}
Miniaturized Internet-of-Things (IoT) systems have unique advantages over larger devices due to their ability to provide contextual information at the data source. They can be particularly advantageous in healthcare scenarios, where medically implanted devices need to be non-obtrusive for long-term \textit{in vivo} monitoring \cite{imd}, and in environmental and animal monitoring applications, where larger instrumentation could disturb the target species or be unable to capture hyper-local data \cite{snail}\cite{beetle}. 

Embedded vision systems have the potential to transform application spaces, such as industrial monitoring particularly in remote or hazardous locations. In spite of these benefits, size-constrained sensing systems pose several fundamental challenges, all of which are exacerbated by using a data-intensive and power-hungry visual sensing modality. The following technical challenges unique to mm-scale systems are tackled in our proposed system: 

 \begin{figure}[b]
\centering
\includegraphics[width=0.45\textwidth]{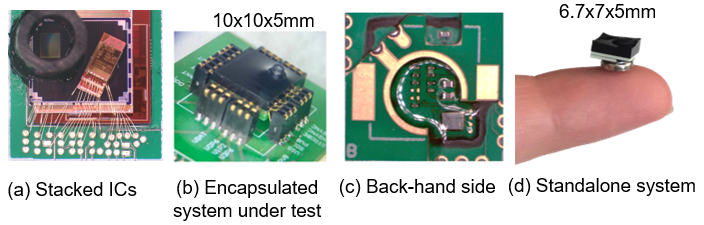}
\caption{\textbf{Proposed embedded imaging system}}
\label{imager}
\end{figure}

\begin{figure*}[h!]
\centering
\includegraphics[width=0.88\textwidth]{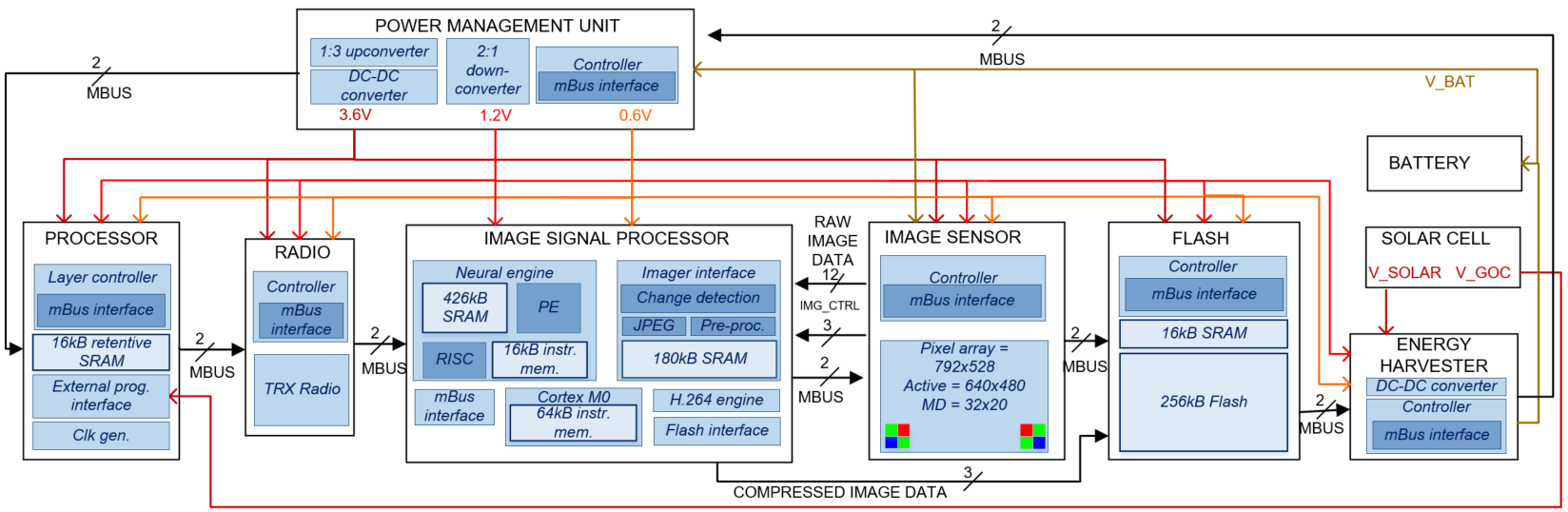}
\caption{\textbf{System data and power interconnect diagram}}
\label{block diagram}
\end{figure*}

\begin{figure}[h]
\centering
\includegraphics[width=0.43\textwidth]{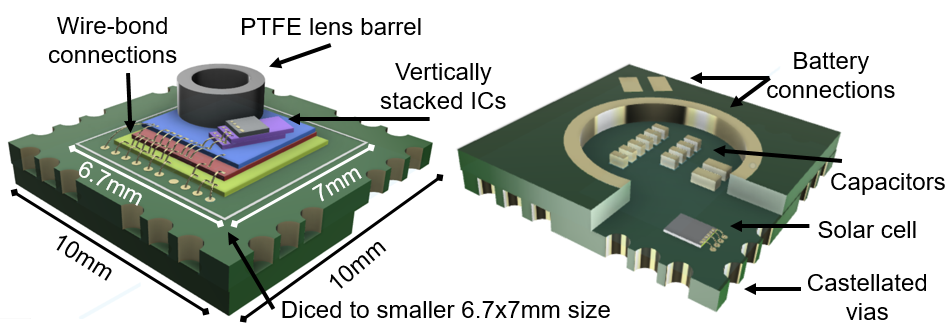}
\caption{\textbf{Imager system cross-section}}
\label{crosssec1}
\end{figure}

\textbf{1) Millimeter (mm)-scale size: }Designing a highly area constrained system requires integrating ultra-low power (ULP) ICs and assembling them to minimize system size. A mm-scale battery is needed to maintain small system dimension, imposing tight energy and peak power constraints.

\textbf{2) Constrained energy budget and peak current: } Millimeter-scale batteries have limited capacities (1-50 mAh) and stringent peak current constraints whereas always-on image sensors have  high power consumption, particularly compared to other sensing modalities such as audio. Imaging systems often include computation- and data-intensive digital signal processing and analytic algorithms to perform advanced tasks such as image compression and object/event detection. Deep neural networks (DNNs) can enhance the accuracy of these algorithms but often with the expense of higher number of operations and stored parameters.

\textbf{3) Limited data storage: }Unlike other scalar sensing modalities, raw image data has a significant memory footprint of 3.7Mb per VGA (640$\times$480$\times$3 pixel) frame. A mm-scale system will have a limited amount of on- and off-chip memory available, limiting the storage of image data and the complexity of the algorithms used.

\textbf{4) Costly wireless communication: } ULP wireless communication imposes an overall high energy cost due to low data rates ($\leq$1Mbps), making it inefficient to offload large volumes of data. Commercial off-the-shelf Bluetooth low energy wireless solutions provide a higher data rate but require a reference crystal and matching antenna with relatively high active power consumption of $\approx$10 mW, placing it outside our system constraints.

\textbf{5) Non-idealities from mm-scale lens and ULP front-end: }Though using a mm-scale lens and ULP imaging front-end enable unobtrusive long-term operation, the resulting images are susceptible to color and geometric distortions. DNN algorithms that have been trained on high quality image datasets suffer significant performance degradation when applied to mm-scale IoT systems. 

To address these challenges, work has been done on low-power image sensors \cite{ulpimagesensor}\cite{imagesensorandacc}, ULP processors \cite{iotsoc}\cite{realtimeimageproc} and mixed-signal vision ICs \cite{imagersoc}\cite{senseandcompute}, low-power wireless communication \cite{crystallesstx}\cite{ulpblerx}, efficient neural accelerators \cite{inmemaccelerator}\cite{realtimegr}, and optimizing machine learning algorithms for embedded applications \cite{embeddedml}. However, limited prior work has combined these approaches into a functional system. Prior demonstrated systems \cite{flyrobot}\cite{beetle}\cite{wirelesssys}\cite{mmimage}\cite{continuouscv} either do not incorporate edge intelligence or do not meet the area/power-constraints that motivate this paper.

This paper proposes and demonstrates a fully integrated mm-scale imaging system with deep-learning and image processing capabilities, which is the first of its kind. By using a combination of techniques specific to ULP tiny-IoT systems that tackle the above challenges, we achieve a lifetime of 7 days without recharging the battery and an overall average power consumption of 49.6$\upmu$W. Long-term sustained operation may be achieved by using the system's energy harvesting IC to recharge the battery.
\label{Introduction}
\vspace{-2.5mm}
\section{System Description}
\vspace{-1mm}
\subsection{Stacked custom ICs}
The system uses un-packaged and thinned (150$\upmu$m) ULP ICs from a family of custom chips that are prior-designed to be vertically stacked and interconnected by attaching wire-bonds between IC bond pads as shown in Fig. \ref{imager}a. This method increases the number of ICs that can be integrated in the same footprint in comparison to traditional planar 2D chip-to-chip connection on a PCB, allowing the system to achieve a 6.7$\times$7$\times$5mm size and 460mg weight (battery included). The integrated system (Fig. \ref{imager} and Fig. \ref{crosssec1}) consists of:

\textbf{1) A base-layer:} It integrates the following into a single IC die to reduce stacking height and wire bonding complexity.
\vspace{-1.6mm}
\begin{itemize}
    \item A master controller, containing a Cortex-M0 microprocessor and 16kB SRAM, that is designed for ULP operation, provides the clock and acts as a bus mediator for the open-source ULP bus protocol \textit{mBus} \cite{mbus}, used for inter-layer communication.
    \item A power management unit that generates 0.6V, 1.2V and 3.6V domains from a single battery voltage in the range of 0.9-4V, and maintains high conversion efficiency under loads ranging from nW to hundreds of $\upmu$W \cite{pmu}. 
    \item A radio IC that uses energy-efficient sparse pulse position modulation consuming <70$\upmu$W of average active power \cite{mrr}. It is connected to a metal trace loop antenna integrated within the PCB, and uses a carrier frequency of 1.2GHz and sampling frequency of 240kHz generated by an on-chip oscillator without a crystal reference.
\end{itemize}
\vspace{-1.6mm}
\textbf{2) A ULP image sensor layer: } It supports motion-triggered 12-bit VGA image capture and also near-pixel motion detection on a sub-sampled 32$\times$20 pixel frame at a maximum rate of 170 fps  \cite{vimjournal}.

\textbf{3) A ULP image signal processing (ISP) layer: } It is a revised version of \cite{isp}, that performs on-the-fly JPEG (de)compression, optical-black pixel calibration, de-Bayering, RGB-to-YUV conversion, and scene change detection when an image is streamed in. It has a H.264 compression engine and 180kB of SRAM to store compressed frames on-chip. A neural engine (NE) with a peak efficiency of 1.5 TOPS/W enables DNN-based frame analysis, and has 426kB SRAM and a custom RISC-like processor. An internal ARM Cortex-M0 microprocessor orchestrates these components. The ISP's memory banks have separately-controllable power-gating switches to reduce leakage energy for non-retentive data.

\textbf{4) Two stacked, custom ultra-low-leakage flash layers:} They provide a total of 16kB SRAM storage and 256kB flash storage with 11pJ/bit read energy and pW sleep power \cite{flash}.

\textbf{5) An energy harvester layer:} It is designed to be supplied by a low-voltage source that has a configurable conversion ratio to increase harvesting efficiency for a range of input voltages \cite{hrv}.

\textbf{6) A solar cell layer:} It is for energy harvesting and a global optical communication \cite{sol}.

\textbf{7) A 3V 3.4mAh 5.8$\times$1.8mm rechargeable lithium battery:} It weighs 130mg with 150$\upmu$A maximum continuous discharge current and 10$\upmu$A standard discharge current.

\textbf{8)A polytetrafluoroethylene (PTFE) tube: } It is used as a lens barrel, housing a medical endoscopic lens. 

Layers are connected in a daisy-chain configuration and forward \textit{mBus} transactions between a transmitting and a receiving layers as shown in Fig. \ref{block diagram}. The ISP's image interface connects directly to the image sensor IC pads for raw data and synchronization signals. The ISP's flash interface allows it to stream (de)compressed image data directly to flash. The power management unit  generates 3 power domains that are distributed throughout the system. The battery is recharged by the harvester connected to the solar cell layer.

\vspace{-3mm}
\subsection{System design and integration}
 The system uses two 4 layer 10$\times$10$\times$0.8mm PCBs. The front of the first PCB is used for wire-bonding the stacked ICs whereas passive components and the solar cell layer are placed on the back (Fig. \ref{imager}c). The second PCB provides spacing between passive components and the battery so these can be placed above each other.

Since code development and debug on miniature systems is extremely difficult due to limited physical access, we include castellated vias on the outer rims of the system to expose internal signals that connect to compression contacts in our test setup (Fig.\ref{imager}b). Fig. \ref{crosssec1} shows a white outline where the system can be diced a second time into an even smaller form-factor of 6.7$\times$7$\times$5mm (Fig. \ref{imager}d). With this smaller size, the system loses all exposed probe contacts for debugging and monitoring, thus must be wirelessly programmed through an optical communication interface using the solar cell.

Fig. \ref{imager} shows a completed assembly, where the stacked ICs have been encapsulated in black epoxy to block out light due to ULP IC's light-sensitivity. The solar cell is encapsulated with clear epoxy to be exposed to light. The image sensor is also left exposed by attaching the PTFE lens barrel over it prior to encapsulation.

\label{System Description}
\vspace{-2.5mm}
\section{System Demonstration}

\begin{figure}[t]
\centering
\includegraphics[width=0.43\textwidth]{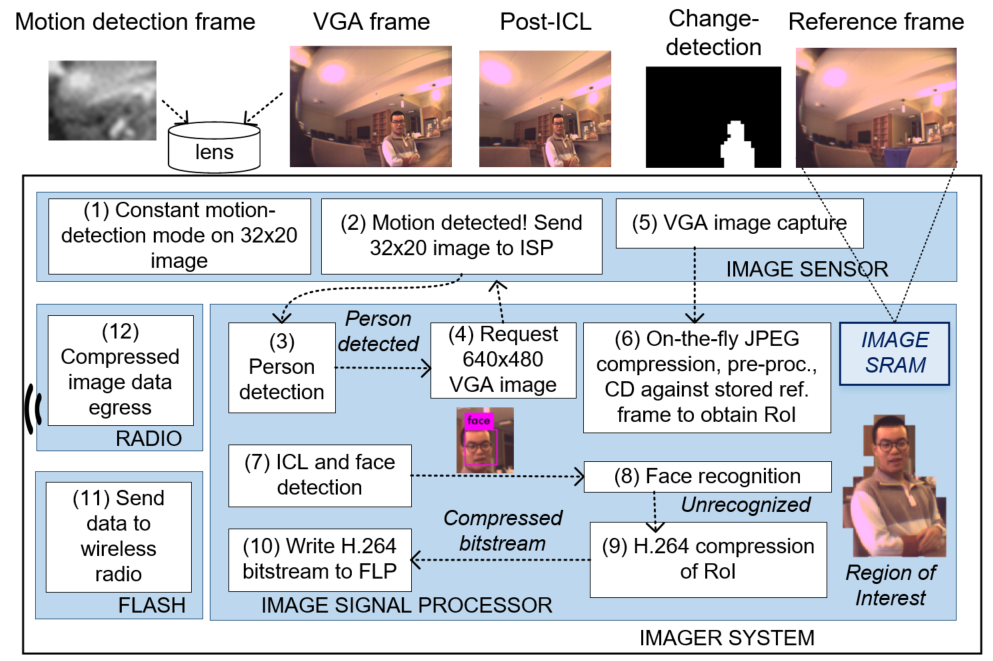}
\caption{\textbf{Intruder detection data flow}}
\label{data flow}
\vspace{-4.5mm}
\end{figure}

The system is fully programmable for various applications. As a driving application example, we demonstrate the system for home-surveillance as depicted in Fig. \ref{data flow}. In the demonstrated scenario, the image sensor captures highly sub-sampled (32$\times$20$\times$1 pixel) frames at a rate of 1fps and performs motion detection on-chip. When motion is detected, the ISP receives the sub-sampled frame and performs DNN-based person detection. If a person is not detected, the system returns to continuous motion-detection mode. Otherwise, a VGA frame is taken by the image sensor and streamed to the ISP, which performs image pre-processing, JPEG compression and change detection on the scene by comparing it to a stored reference frame to obtain a pruned (non-rectangular) region of interest (RoI).

Then, the VGA image is corrected for geometric distortion and resized for YOLO-Lite-based face detection. It is a one-shot object detection and localization DNN based on \cite{yolo}, widely researched for embedded vision systems, and we customized it to satisfy the system constraints. If a face is not detected,  the system returns to the continuous motion-detection mode. Otherwise, the scene is analyzed by performing DNN-based face recognition. The image RoI is H.264 compressed and stored in the system's flash layer only when a face is not recognized (i.e., a possible intruder) for optional wireless transmission to a custom gateway. The VGA frame containing an intruder is reconstructed offline.
\label{intruderdetection}
\vspace{-2.5mm}
\section{Energy Minimization Techniques}
\begin{figure}[t]
\includegraphics[width=0.43\textwidth]{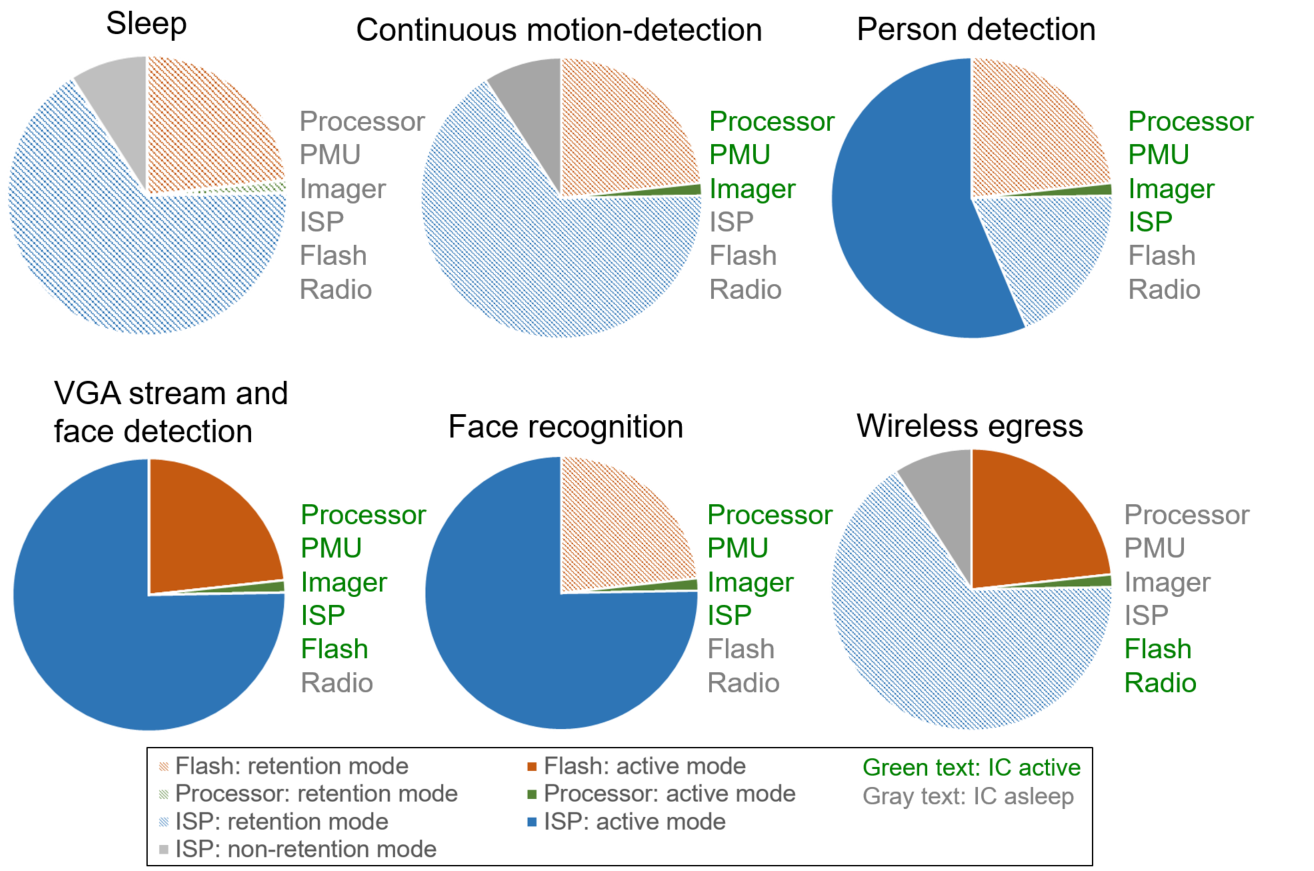}
\caption{\textbf{Layer power and memory state per behavior}}
\label{pmu part}
\end{figure}

The chosen mm-scale battery has a limited capacity, making power management crucial to achieve a standalone system with a reasonable operating lifetime. This is particularly challenging for always-on imaging systems, where power consumption is dominated by the continuously running image sensor. Performing DNN-based analysis is costly due to both the computation energy and the mega-bytes of DNN parameters stored on-/off-chip. However, intelligent DNN-based classifications on the edge is crucial in order to reduce wireless data offloading, which consumes $\sim$15nJ/b and can dominate the system's energy consumption for large data volumes if not trimmed by DNN-based edge processing.

\vspace{-2mm}
\subsection{Dynamic energy-efficient modes}
Most ICs in the proposed system are duty-cycled to optimize the system's power consumption for the chosen application. Fig.\ref{pmu part} shows the power breakdown of each mode. For instance, flash ICs are set to sleep during motion monitoring, image capture, and DNN-based scene analysis, consuming only 0.003$\upmu$W. ISP allows for fine-grained software control of its SRAM banks, which can be in low-power retention mode without loss of data. The NE SRAM is partitioned into 7 variably-sized blocks that can be selectively power-gated to reduce leakage power consumption according to DNN storage requirements. The H.264 engine and image interface memory also have this capability. For example, when the system is in continuous motion-detection mode, the 430kB ISP memory can be power-gated, reducing the leakage power by 1.8$\times$. The power management unit is adjusted at each mode to maximize the efficiency for the dynamic load by modifying current draw, frequency control and up/down conversion ratio, achieving an average efficiency of 64\%. 

\begin{figure*}[t]
\centering
\includegraphics[width=0.76\textwidth]{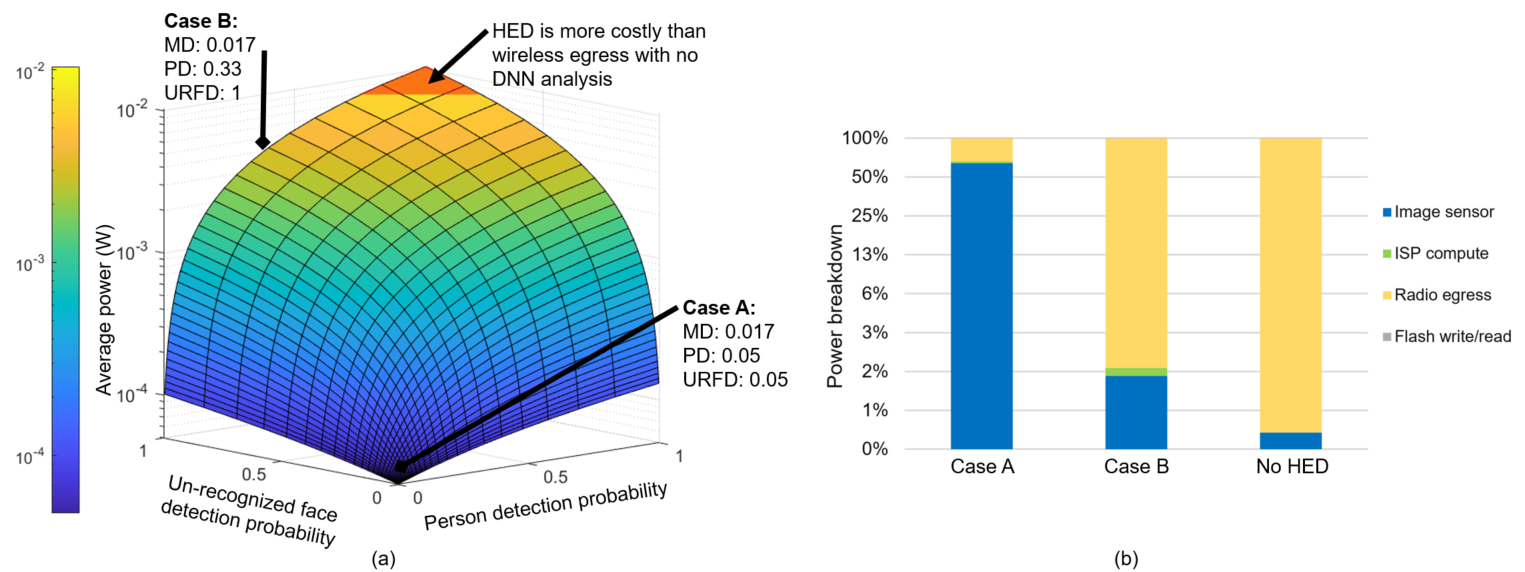}
\caption{\textbf{Hierarchical event detection effect on average system power}}
\label{HED graphs}
\end{figure*}

\vspace{-3mm}
\subsection{Hierarchical event detection}
\label{Hierarchical ED}
A hierarchical event detection (HED) algorithm is employed to prune out irrelevant events that would otherwise wastefully consume energy, particularly when offloading data without determining its value to the user. This approach uses efficient motion detection in the imager front-end and lower complexity DNN-analysis in the ISP back-end to reduce the frequency at which costly VGA images and higher-complexity DNN algorithms are performed, significantly decreasing the system's average power/energy.

While the image sensor IC is in motion-detection mode, which is performed on the sensor without requiring external processing or storage, other layers are set to sleep so that the system's power consumption of 48.8$\upmu$W is dominated by the image sensor. If motion is detected, the system wakes up the ISP, which performs person detection on a sub-sampled frame, consuming 39.6$\upmu$J. If a person is detected, the image sensor takes a VGA frame, consuming 349$\upmu$J and 14.9$\upmu$J to stream the frame into the ISP so that it can perform face detection (643$\upmu$J). If a face is identified, face recognition is performed consuming 622$\upmu$J. Only when an unregistered face is detected, the frame is stored in flash and offloaded wirelessly.

Fig. \ref{HED graphs}a shows the average system power for the above scenario, when the image sensor's motion detection event rate is fixed to once per minute while the person detection (PD) and the un-registered face detection (URFD) event probabilities are varied. The URFD rate combines both face detection and face recognition, and it is assumed that data written to flash and sent wirelessly is uncompressed. This analysis motivates the use of HED for the chosen application, whereas the red top corner demarcates the PD and URFD probabilities where performing HED yields no power savings. In all other scenarios, HED enables significant (up to $\approx100\times$) power reduction especially when both URFD and PD rates are low. Fig. \ref{HED graphs}b shows the system power breakdown for 3 event cases. When the total event rate is low (`Case A'), the system power is dominated by the image sensor, while the wireless power dominates as the event rate increases. Notably, the ISP computation power for DNNs is always insignificant compared to the image sensor or radio, showing the benefit of HED in reducing overall power.

\label{Algorithms_energy}
\vspace{-2.5mm}
\section{Data Compression Techniques}
\label{Data compression}
The proposed mm-scale imaging system has limited IC area for flash memory and on-chip SRAM, making it critical to employ aggressive data compression techniques for DNN parameters and images. 
\vspace{-4mm}
\subsection{DNN parameter compression}

High compression of up to 1.5bit/weight is achieved for DNNs by combining weight pruning, non-uniform quantization, Huffman encoding of quantized weights for convolutional layers, and index-based encoding for sparse fully-connected layers. The compressed sizes for person detection, face detection, and face recognition networks are 85kB, 112kB, and 107kB, respectively. %which reduce SRAM leakage and inter-chip data movement energy. 
Up to 1.75M compressed weights can be stored in the ISP NE on-chip SRAM, which are on-the-fly decompressed when moved to NE local SRAM buffers. Once weights are loaded into the SRAM buffers, they are heavily reused for convolution on large input activations so that the energy overhead of decompressing weights is negligible.

The proposed HED requires multiple DNNs but parameters for less frequently executed DNNs can be stored off-chip in flash to eliminate on-chip SRAM leakage power. This off-chip parameter loading approach, however, increases execution time and incurs energy overhead from inter-chip data movement. We first analyze the event frequency at which this on- vs. off-chip parameter storage trade-off is advantageous for each DNN, and then program each DNN based on the assumption of expected event frequency to minimize the average power consumption.

\vspace{-2mm}
\subsection{Image compression}
We use a combination of several compression methods to minimize the data footprint and reduce wireless transmission cost.

\vspace{-2mm}
\subsubsection{JPEG and H.264 compression}

The ISP performs on-the-fly JPEG compression on frames streamed in by the image sensor on a macro-block (MCB, 16$\times$16 pixel) level, reducing the memory footprint of storing images on-chip and the SRAM leakage power by 11$\times$ and only consuming 14.9$\upmu$J. As a result, both a reference and current frame can be stored on-chip in 180kB SRAM rather than off-chip, which otherwise would incur energy overheads of 407$\upmu$J/frame. We leverage the ISP's customized H.264 intra-frame compression engine to reduce the memory footprint of a VGA frame by 23$\times$. We only apply H.264 to the image containing an unregistered face instead of all incoming VGA frames due to the higher latency (than JPEG) that cannot meet the image sensor interface throughput for on-the-fly compression. Though H.264 compression incurs 138\% more processing energy than JPEG, at a system-level it reduces the flash write energy from 36.9$\upmu$J to 17.4$\upmu$J and the wireless egress energy from 37.7mJ to 17.8mJ as it attains a higher compression rate.

\vspace{-2mm}
\subsubsection{Change detection engine}
\label{CD}

Change detection between a new frame and reference frame stored on-chip can be performed on-the-fly by the ISP. Each MCB of an input frame is encoded into a 64b pattern vector and compared to the stored image vector to create a 40$\times$30 change detection map representing a non-rectangular RoI. We use a customized H.264 algorithm, which removes MCB inter-dependence, to compress only the non-rectangular RoI while unchanged MCBs are pruned. This achieves 135$\times$ compression compared to a VGA frame while consuming only 2.8$\upmu$J (assuming 86\% scene pruning observed in our test dataset). At a system-level, combining change detection and H.264 compression reduces the flash write energy from 17.4$\upmu$J to 3$\upmu$J, the wireless egress energy from 17.8mJ to 3.1mJ, and allows for 74 images to be stored in flash.

\vspace{-2mm}
\subsubsection{Off-system image reconstruction}

To reduce the wireless data egress, we reconstruct the current frame offline by only sending the change detection map (1.2kb) and the H.264 compressed RoI of the current frame (20-30kb) to a custom gateway. We assume that the imager system is deployed in a stationary location, so the system and gateway share the same reference image, which is periodically reprogrammed. A scene is reconstructed by decompressing the unchanged MCBs from the reference image and the changed MCBs from the current frame, reducing the egress data by 130$\times$.

\label{Algorithms_data}
\vspace{-3mm}
\section{Image correction for \lowercase{mm}-scale Lens and ULP Front-Ends}
Most cameras have $\sim$60$^{\circ}$ field of view (FOV), much less than a human. To achieve a larger FOV and capture more information for wide-view scene analysis, our system adopts a wide-angle lens. However, using this lens and a ULP image sensor that lacks sophisticated image pre-processing results in geometric and color distortion, negatively affecting the performance of deep learning perceptual tasks, as was demonstrated when deploying our system.

\begin{figure}[t]
\centering
\includegraphics[width=0.45\textwidth]{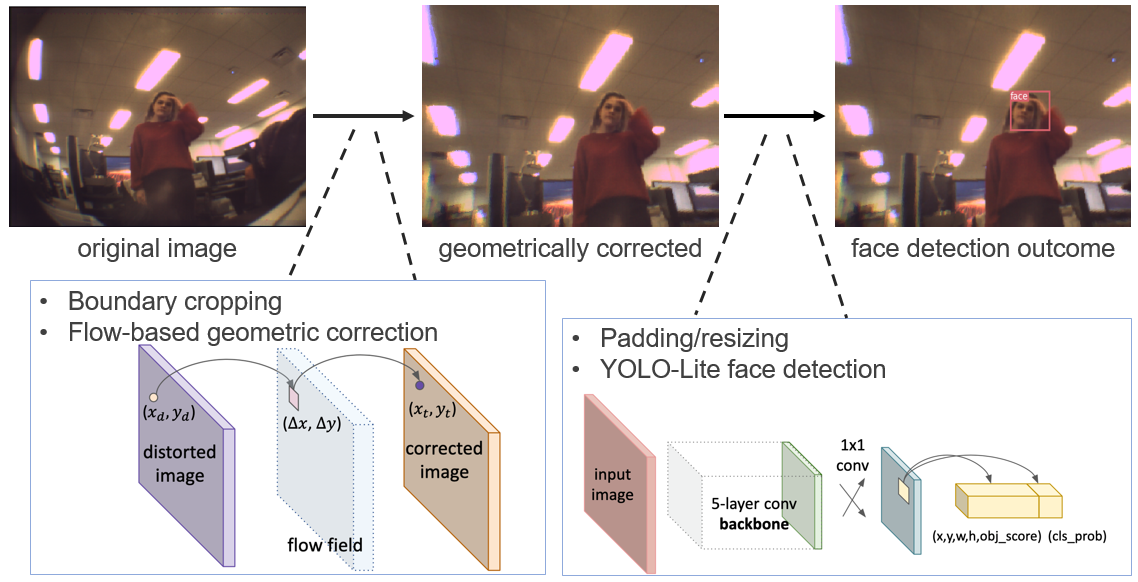}
\caption{\textbf{Pipeline for ICL and YOLO face detection}}
\label{IENN_pipeline}
\end{figure}

We propose image-correcting layers (ICL) that can be executed on the ISP NE using supported instructions, such as matrix multiplication and convolution, to compensate for artefacts and distortions (Fig. \ref{IENN_pipeline}). Since ICL is based on DNN-compatible instructions, it does not require a dedicated hardware accelerator and can be generalized for other constrained systems.

\vspace{-4mm}

\subsection{Distortion modeling and compensation}
The fisheye-like lens distortion is modeled by capturing checkerboard images using our system. Geometric distortion correction can be expressed by a mapping between pixels in the distorted image $I_d$ and the pixels in the correct image $I_t$: $(x_d, y_d) \rightarrow (x_t, y_t) \;\;\; \forall (x_d, y_d) \in I_d$. Various geometric distortion correction algorithms have been extensively studied and implemented \cite{distortion-1}\cite{distortion-3}. However, for real-time processing, a majority of them require task-specific hardware accelerators which are not available on our system. 

\begin{table}[h]
    \centering
    \begin{tabular}{|c|c|c|c|}
        \hline
        SCHEME & NETWORK SPARSITY & AVG. IOU & AVG. RECALL \\
        \hline
        \hline
        1 & 0\% & 0.7570   & 0.9260 \\
        \hline
        2 & 0\% & 0.5769  & 0.7213 \\
        \hline
        3 & 0\% & 0.6648 & 0.7885 \\
        \hline
        4 & 0\% & 0.7453 & 0.8947 \\
        \hline
        5 & 50\% & 0.7172 & 0.8449\\
        \hline
        6 & 50\% & 0.6254 & 0.7647 \\
        \hline
    \end{tabular}
    \caption{\textbf{Face Detection Evaluation Results}}
    \label{tab:IE}
\end{table}

Instead, we construct a $640\times480\times2$ flow field matrix (FFM) that represents vertical and horizontal pixel-displacement vectors based on the offline constructed fisheye distortion model. This ICL can be performed by either the ISP Cortex-M0 or NE by representing the FFM as a sparse fully connected layer. This is advantageous compared to both conventional and learning-based distortion correction methods since it only requires a relatively small correction matrix, which has low memory and computational complexity. 

Color correction is performed using a color correction matrix (CCM) that projects the \emph{RGB} pixel values in the distorted image to the target \emph{RGB} values at the same location in the corrected image, expressed by a linear equation $[R^t_{ij},G^t_{ij},B^t_{ij}] = [R^d_{ij},G^d_{ij},B^d_{ij}] \mathbf{A} $ for pixel $(i,j)$ where $\mathbf{A}$ denotes the CCM. Our CCM is obtained by capturing images of an eSFR (edge spatial frequency response) chart using the proposed system. The color restoration ICL is implicitly accomplished with a linear layer in the custom face detection network. Training data images are modified using the color distortion matrix to emulate the system and implicitly learn the CCM.

\begin{table*}[t!]
    \centering
    \begin{adjustbox}{width=1\textwidth}
    \begin{tabular}{|c|c|c|c|c|c|c|}
        \hline
         METHOD & SIZE & PROCESSING ENERGY & FLASH WRITE ENERGY & EGRESS ENERGY & EGRESS TIME & VGA CAPACITY \\
        \hline
        \hline
        Raw VGA Image & 3.7Mb & 0 & 407$\upmu$J & 414mJ & 2 hours & 0 \\
        \hline
        JPEG  & 335kb & 11.9$\upmu$J & 36.9$\upmu$J & 37.7mJ &  11.6 min & 1 \\
        \hline
        H.264 (QF=20) & 158.2kb & 28.3$\upmu$J & 17.4$\upmu$J & 17.8mJ & 5.5 min & 12\\
        \hline
        H.264 + CD (QF=20) & 27.4kb & 14.8$\upmu$J & 3$\upmu$J & 3.1mJ & 57s & 74\\
        \hline
    \end{tabular}
    \end{adjustbox}
    \caption{\textbf{Compression gains on overall system operation}}
    \label{tab:compression h264}
\end{table*}

\vspace{-3mm}
\subsection{Experiments on ICLs for Face Detection}

The proposed ICLs were tested using images in the COCO-faces dataset, a dataset with annotated human faces sourced from COCO 2017 \cite{coco-dataset}, and on images captured by our system. We first pre-train the customized YOLO-Lite network without distortion to speed up convergence, then apply the distortion models to the training/validation dataset, and include the pre-determined geometric distortion correction ICL. Finally, we fine-tune the YOLO model for color correction. 

The performance of our proposed YOLO-Lite face detection network with ICLs is quantified by evaluating the average intersection over union (IoU) and average recall. In Table \ref{tab:IE}, we evaluate six schemes: 1) original images, 2) images with modeled imager geometric and color distortions, 3) geometric distortion correction, 4) additionally fine-tuning the face detection network using the color distorted training dataset, 5) additionally applying 50\% network pruning, 8-bit fixed point quantization, and parameter compression to the final network (necessary for on-chip parameter storage), and 6) testing scheme 5 on a 112$\times$112 pixel gray-scale input image.

The results show a substantial performance drop after applying the modeled distortion to the original image data. Both performance measures are improved to a level close to the original when lens distortion and color restoration ICLs are applied. We experimented on the impact of using a smaller input gray-scale image, together with down-graded intermediate feature maps (case 6 in Table \ref{tab:IE}). The results demonstrate that our proposed image correction and face detection pipeline give sensible performance in this case, showing that it can be extended to other highly-constrained systems.

\label{image enhancer}
\vspace{-3mm}
\section{Integrated System Evaluation}

We have integrated a mm-scale, standalone intelligent imaging system (Fig.\ref{imager}) for an intruder detection demonstration while employing the methods discussed in Sections 2,4-6. Table \ref{tab:compression h264} outlines the gains of each compression method relative to a raw VGA image, which yield proportional energy savings when writing the image data to flash and transmitting the data via the ULP wireless transmitter. Using the combination of H.264 and change detection-based compression significantly decreases the egress energy to 3.1mJ and the egress time to 5 minutes, making the system more practical for a real-life scenario. This final method extends the lifetime when performing continuous motion-detection, VGA capture, flash storage and wireless egress by 13$\times$ without battery recharging.

\begin{figure}[h]
\centering
\includegraphics[width=0.43\textwidth]{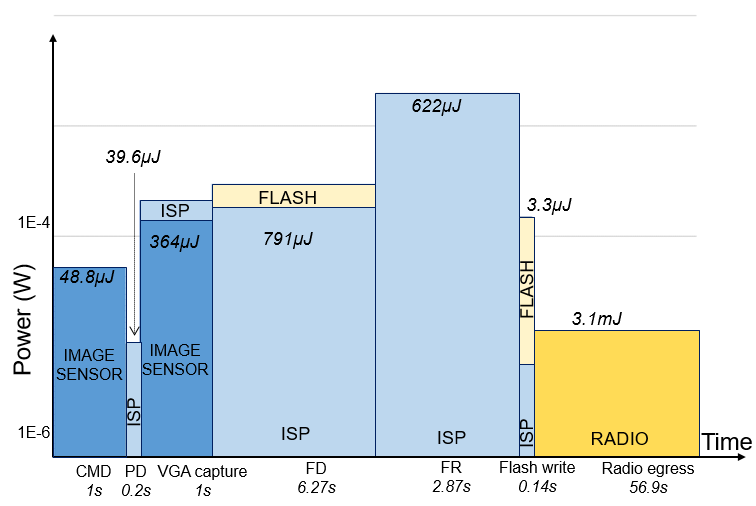}
\caption{\textbf{System energy per behavior and stage}}
\label{behavior energy}
\end{figure}

Fig. \ref{behavior energy} shows the intruder detection scenario flow using HED, including the energy consumption and execution time of each system behavior. The ISP runs at a relative low clock frequency of 240kHz (programmable) to maintain low power. This processing performance (latency of each step in Fig. \ref{behavior energy}) is acceptable for our target scenario where the event of interest is infrequently triggered. Fig. \ref{data flow} shows examples of actual images processed by the system through the proposed multi-DNN and image compression data path.

The radio's link distance and data rate can be adjusted by modifying the number of repeated pulses, the pulse width and separation, and the decap recharging time. Two different modes were selected and measured in a line-of-sight indoor environment. The long distance mode achieves a link distance of 20m, a data rate of 256bps, and an egress time for a compressed image of 106s. The short distance mode  has a shorter distance of 15m, but it achieves a higher data rate of 480bps and shorter transmit time of 56s.

 For system lifetime analysis, we assume that a motion-triggered sub-sampled frame is sent to the ISP every 20 minutes to perform person detection. When HED is used (assuming event ratios of 1:2, 1:2 and 1:5 for PD, FD and FR), the overall average power consumption is 49.6$\upmu$W for an expected battery lifetime of 7 days, which can be extended by trading off larger system size for a higher capacity battery. The total power consumption of the system when the ISP layer is not retaining any image or DNN data is 7.35$\upmu$W, giving a total shelf-life of 48 days without recharging. In this (re)charging mode, the harvester and solar layers can provide $\sim$10$\upmu$W to recharge the battery under 10klux lighting conditions to extend the total shelf-life until system (re)deployment.

\label{System evaluation}
\vspace{-2.5mm}
\section{Conclusion}

This paper proposes a novel mm-scale, standalone system with deep-learning and image processing for intruder detection, with an average power consumption of 49.6$\upmu$W and expected lifetime of 7 days without recharging. It achieves a miniature size by vertically stacking ULP custom ICs, and satisfies a tight memory and energy budget through a combination of data/energy management methods. We present image-enhancement layers to improve DNN performance in constrained embedded systems. Comprehensive system-level analysis was conducted to quantify the gain of proposed techniques, which can be generalized to other ULP systems. Having demonstrated a tiny-IoT intelligent imaging system, analysis through a sociotechnical and ethical lens is an essential next step; we invite future work on topics such as security and privacy.

\label{Conclusion}

\vspace{-2.5mm}
\begin{acks}
The authors would like to thank Sony Semiconductor Solutions Corp./Sony electronics Inc. for supporting this work.
\end{acks}

\bibliographystyle{ACM-Reference-Format}
\bibliography{acmart.bib}


\begin{thebibliography}{30}

%%% the Web address.

\ifx \showCODEN    \undefined \def \showCODEN     #1{\unskip}     \fi
\ifx \showDOI      \undefined \def \showDOI       #1{#1}\fi
\ifx \showISBNx    \undefined \def \showISBNx     #1{\unskip}     \fi
\ifx \showISBNxiii \undefined \def \showISBNxiii  #1{\unskip}     \fi
\ifx \showISSN     \undefined \def \showISSN      #1{\unskip}     \fi
\ifx \showLCCN     \undefined \def \showLCCN      #1{\unskip}     \fi
\ifx \shownote     \undefined \def \shownote      #1{#1}          \fi
\ifx \showarticletitle \undefined \def \showarticletitle #1{#1}   \fi
\ifx \showURL      \undefined \def \showURL       {\relax}        \fi
% invisible to TeX
\providecommand\bibfield[2]{#2}
\providecommand\bibinfo[2]{#2}
\providecommand\natexlab[1]{#1}
\providecommand\showeprint[2][]{arXiv:#2}

\bibitem[\protect\citeauthoryear{Ahlen, Broszio, and Wassermann}{Ahlen
  et~al\mbox{.}}{2003}]%
        {distortion-1}
\bibfield{author}{\bibinfo{person}{T. Ahlen}, \bibinfo{person}{H. Broszio},
  {and} \bibinfo{person}{I. Wassermann}.} \bibinfo{year}{2003}\natexlab{}.
\newblock \showarticletitle{Robust Line-Based Calibration of Lens Distortion
  from a Single View}.
\newblock \bibinfo{journal}{\emph{Mirage 2003}} (\bibinfo{date}{05}
  \bibinfo{year}{2003}).
\newblock


\bibitem[\protect\citeauthoryear{Ajani, Imoize, and Atayero}{Ajani
  et~al\mbox{.}}{2021}]%
        {embeddedml}
\bibfield{author}{\bibinfo{person}{T.~S. Ajani}, \bibinfo{person}{A.~L.
  Imoize}, {and} \bibinfo{person}{A.~A. Atayero}.}
  \bibinfo{year}{2021}\natexlab{}.
\newblock \showarticletitle{An Overview of Machine Learning within Embedded and
  Mobile Devices–Optimizations and Applications}.
\newblock \bibinfo{journal}{\emph{Sensors}} \bibinfo{volume}{21},
  \bibinfo{number}{13} (\bibinfo{year}{2021}).
\newblock
\showISSN{1424-8220}
\urldef\tempurl%
\url{https://doi.org/10.3390/s21134412}
\showDOI{\tempurl}


\bibitem[\protect\citeauthoryear{{An}, {Venkatesan}, {Schiferl}, {Wesley},
  {Zhang}, {Wang}, {Choo}, {Liu}, {Liu}, {Li}, {Zhong}, {Gong}, {Blaauw},
  {Dreslinski}, {Sylvester}, and {Kim}}{{An} et~al\mbox{.}}{2020}]%
        {isp}
\bibfield{author}{\bibinfo{person}{H. {An}}, \bibinfo{person}{S. {Venkatesan}},
  \bibinfo{person}{S. {Schiferl}}, \bibinfo{person}{T. {Wesley}},
  \bibinfo{person}{Q. {Zhang}}, \bibinfo{person}{J. {Wang}},
  \bibinfo{person}{K. {Choo}}, \bibinfo{person}{S. {Liu}}, \bibinfo{person}{B.
  {Liu}}, \bibinfo{person}{Z. {Li}}, \bibinfo{person}{H. {Zhong}},
  \bibinfo{person}{L. {Gong}}, \bibinfo{person}{D. {Blaauw}},
  \bibinfo{person}{R. {Dreslinski}}, \bibinfo{person}{D. {Sylvester}}, {and}
  \bibinfo{person}{H.~S. {Kim}}.} \bibinfo{year}{2020}\natexlab{}.
\newblock \showarticletitle{A 170$\upmu$W Image Signal Processor Enabling
  Hierarchical Image Recognition for Intelligence at the Edge}. In
  \bibinfo{booktitle}{\emph{2020 IEEE Symposium on VLSI Circuits}}.
  \bibinfo{pages}{1--2}.
\newblock


\bibitem[\protect\citeauthoryear{Balasubramanian, Chukewad, James, Barrows, and
  Fuller}{Balasubramanian et~al\mbox{.}}{2018}]%
        {flyrobot}
\bibfield{author}{\bibinfo{person}{S. Balasubramanian}, \bibinfo{person}{Y.~M.
  Chukewad}, \bibinfo{person}{J.~M. James}, \bibinfo{person}{G.~L. Barrows},
  {and} \bibinfo{person}{S.~B. Fuller}.} \bibinfo{year}{2018}\natexlab{}.
\newblock \showarticletitle{An Insect-Sized Robot That Uses a Custom-Built
  Onboard Camera and a Neural Network to Classify and Respond to Visual Input}.
  In \bibinfo{booktitle}{\emph{2018 7th IEEE International Conference on
  Biomedical Robotics and Biomechatronics (Biorob)}}.
  \bibinfo{pages}{1297--1302}.
\newblock
\urldef\tempurl%
\url{https://doi.org/10.1109/BIOROB.2018.8488007}
\showDOI{\tempurl}


\bibitem[\protect\citeauthoryear{Bick, Lee, Coote, Haponski, and Blaauw}{Bick
  et~al\mbox{.}}{2021}]%
        {snail}
\bibfield{author}{\bibinfo{person}{C.S. Bick}, \bibinfo{person}{I. Lee},
  \bibinfo{person}{T. Coote}, \bibinfo{person}{A.E. Haponski}, {and}
  \bibinfo{person}{D. Blaauw}.} \bibinfo{year}{2021}\natexlab{}.
\newblock \showarticletitle{Millimeter-sized smart sensors reveal that a solar
  refuge protects tree snail \textit{Partula hyalina} from extirpation}. In
  \bibinfo{booktitle}{\emph{Commun Biol}}, Vol.~\bibinfo{volume}{4}.
\newblock
\urldef\tempurl%
\url{https://doi.org/10.1038/s42003-021-02124-y}
\showDOI{\tempurl}


\bibitem[\protect\citeauthoryear{Chen, Chiotellis, Chuo, Pfeiffer, Shi,
  Dreslinski, Grbic, Mudge, Wentzloff, Blaauw, and Kim}{Chen
  et~al\mbox{.}}{2016}]%
        {mrr}
\bibfield{author}{\bibinfo{person}{Y. Chen}, \bibinfo{person}{N. Chiotellis},
  \bibinfo{person}{L. Chuo}, \bibinfo{person}{C. Pfeiffer}, \bibinfo{person}{Y.
  Shi}, \bibinfo{person}{R.~G. Dreslinski}, \bibinfo{person}{A. Grbic},
  \bibinfo{person}{T. Mudge}, \bibinfo{person}{D.~D. Wentzloff},
  \bibinfo{person}{D. Blaauw}, {and} \bibinfo{person}{H. Kim}.}
  \bibinfo{year}{2016}\natexlab{}.
\newblock \showarticletitle{Energy-Autonomous Wireless Communication for
  Millimeter-Scale Internet-of-Things Sensor Nodes}.
\newblock \bibinfo{journal}{\emph{IEEE Journal on Selected Areas in
  Communications}} \bibinfo{volume}{34}, \bibinfo{number}{12}
  (\bibinfo{year}{2016}), \bibinfo{pages}{3962--3977}.
\newblock
\urldef\tempurl%
\url{https://doi.org/10.1109/JSAC.2016.2612041}
\showDOI{\tempurl}


\bibitem[\protect\citeauthoryear{{Choo}, {Xu}, {Kim}, {Seol}, {Wu},
  {Sylvester}, and {Blaauw}}{{Choo} et~al\mbox{.}}{2019}]%
        {vimjournal}
\bibfield{author}{\bibinfo{person}{K.~D. {Choo}}, \bibinfo{person}{L. {Xu}},
  \bibinfo{person}{Y. {Kim}}, \bibinfo{person}{J. {Seol}}, \bibinfo{person}{X.
  {Wu}}, \bibinfo{person}{D. {Sylvester}}, {and} \bibinfo{person}{D.
  {Blaauw}}.} \bibinfo{year}{2019}\natexlab{}.
\newblock \showarticletitle{Energy-Efficient Motion-Triggered IoT CMOS Image
  Sensor With Capacitor Array-Assisted Charge-Injection SAR ADC}.
\newblock \bibinfo{journal}{\emph{IEEE Journal of Solid-State Circuits}}
  \bibinfo{volume}{54}, \bibinfo{number}{11} (\bibinfo{year}{2019}),
  \bibinfo{pages}{2921--2931}.
\newblock


\bibitem[\protect\citeauthoryear{Dong, Kim, Lee, Choi, Li, Wang, Yang, Y.,
  Dong, Cho, Kim, Chang, Chen, Chih, Blaauw, and Sylvester}{Dong
  et~al\mbox{.}}{2017}]%
        {flash}
\bibfield{author}{\bibinfo{person}{Q. Dong}, \bibinfo{person}{Y. Kim},
  \bibinfo{person}{I. Lee}, \bibinfo{person}{M. Choi}, \bibinfo{person}{Z. Li},
  \bibinfo{person}{J. Wang}, \bibinfo{person}{K. Yang}, \bibinfo{person}{Chen.
  Y.}, \bibinfo{person}{J. Dong}, \bibinfo{person}{M. Cho}, \bibinfo{person}{G.
  Kim}, \bibinfo{person}{W. Chang}, \bibinfo{person}{Y. Chen},
  \bibinfo{person}{Y. Chih}, \bibinfo{person}{D. Blaauw}, {and}
  \bibinfo{person}{D. Sylvester}.} \bibinfo{year}{2017}\natexlab{}.
\newblock \showarticletitle{11.2 A 1Mb embedded NOR flash memory with
  39$\upmu$W program power for mm-scale high-temperature sensor nodes}. In
  \bibinfo{booktitle}{\emph{2017 IEEE International Solid-State Circuits
  Conference (ISSCC)}}. \bibinfo{pages}{198--199}.
\newblock


\bibitem[\protect\citeauthoryear{Huang, Pedoeem, and Chen}{Huang
  et~al\mbox{.}}{2018}]%
        {yolo}
\bibfield{author}{\bibinfo{person}{R. Huang}, \bibinfo{person}{J. Pedoeem},
  {and} \bibinfo{person}{C. Chen}.} \bibinfo{year}{2018}\natexlab{}.
\newblock \showarticletitle{YOLO-LITE: A Real-Time Object Detection Algorithm
  Optimized for Non-GPU Computers}. In \bibinfo{booktitle}{\emph{2018 IEEE
  International Conference on Big Data (Big Data)}}.
  \bibinfo{pages}{2503--2510}.
\newblock
\urldef\tempurl%
\url{https://doi.org/10.1109/BigData.2018.8621865}
\showDOI{\tempurl}


\bibitem[\protect\citeauthoryear{Iyer, Najafi, James, Fuller, and
  Gollakota}{Iyer et~al\mbox{.}}{2020}]%
        {beetle}
\bibfield{author}{\bibinfo{person}{V. Iyer}, \bibinfo{person}{Ali. Najafi},
  \bibinfo{person}{J. James}, \bibinfo{person}{S. Fuller}, {and}
  \bibinfo{person}{S. Gollakota}.} \bibinfo{year}{2020}\natexlab{}.
\newblock \showarticletitle{Wireless steerable vision for live insects and
  insect-scale robots}. In \bibinfo{booktitle}{\emph{Science Robotics}},
  Vol.~\bibinfo{volume}{5}.
\newblock
\urldef\tempurl%
\url{https://doi.org/10.1126/scirobotics.abb0839}
\showDOI{\tempurl}


\bibitem[\protect\citeauthoryear{Ji, Pu, Lim, and Horowitz}{Ji
  et~al\mbox{.}}{2016}]%
        {ulpimagesensor}
\bibfield{author}{\bibinfo{person}{Suyao Ji}, \bibinfo{person}{Jing Pu},
  \bibinfo{person}{Byong~Chan Lim}, {and} \bibinfo{person}{Mark Horowitz}.}
  \bibinfo{year}{2016}\natexlab{}.
\newblock \showarticletitle{A 220pJ/pixel/frame CMOS image sensor with partial
  settling readout architecture}. In \bibinfo{booktitle}{\emph{2016 IEEE
  Symposium on VLSI Circuits (VLSI-Circuits)}}. \bibinfo{pages}{1--2}.
\newblock
\urldef\tempurl%
\url{https://doi.org/10.1109/VLSIC.2016.7573545}
\showDOI{\tempurl}


\bibitem[\protect\citeauthoryear{Jia, Ozatay, Tang, Valavi, Pathak, Lee, and
  Verma}{Jia et~al\mbox{.}}{2021}]%
        {inmemaccelerator}
\bibfield{author}{\bibinfo{person}{H. Jia}, \bibinfo{person}{M. Ozatay},
  \bibinfo{person}{Y. Tang}, \bibinfo{person}{H. Valavi}, \bibinfo{person}{R.
  Pathak}, \bibinfo{person}{J. Lee}, {and} \bibinfo{person}{N. Verma}.}
  \bibinfo{year}{2021}\natexlab{}.
\newblock \showarticletitle{15.1 A Programmable Neural-Network Inference
  Accelerator Based on Scalable In-Memory Computing}. In
  \bibinfo{booktitle}{\emph{2021 IEEE International Solid- State Circuits
  Conference (ISSCC)}}, Vol.~\bibinfo{volume}{64}. \bibinfo{pages}{236--238}.
\newblock
\urldef\tempurl%
\url{https://doi.org/10.1109/ISSCC42613.2021.9365788}
\showDOI{\tempurl}


\bibitem[\protect\citeauthoryear{Josephson, Yang, Zhang, and Katti}{Josephson
  et~al\mbox{.}}{2019}]%
        {wirelesssys}
\bibfield{author}{\bibinfo{person}{C. Josephson}, \bibinfo{person}{L. Yang},
  \bibinfo{person}{P. Zhang}, {and} \bibinfo{person}{S. Katti}.}
  \bibinfo{year}{2019}\natexlab{}.
\newblock \showarticletitle{Wireless Computer Vision Using Commodity Radios}.
  In \bibinfo{booktitle}{\emph{Proceedings of the 18th International Conference
  on Information Processing in Sensor Networks}} (Montreal, Quebec, Canada)
  \emph{(\bibinfo{series}{IPSN '19})}. \bibinfo{publisher}{Association for
  Computing Machinery}, \bibinfo{address}{New York, NY, USA},
  \bibinfo{pages}{229–240}.
\newblock
\showISBNx{9781450362849}
\urldef\tempurl%
\url{https://doi.org/10.1145/3302506.3310403}
\showDOI{\tempurl}


\bibitem[\protect\citeauthoryear{Jung, Gu, Myers, Shim, Jeong, Yang, Choi, Foo,
  Bang, Oh, Sylvester, and Blaauw}{Jung et~al\mbox{.}}{2016}]%
        {pmu}
\bibfield{author}{\bibinfo{person}{W. Jung}, \bibinfo{person}{J. Gu},
  \bibinfo{person}{P.~D. Myers}, \bibinfo{person}{M. Shim}, \bibinfo{person}{S.
  Jeong}, \bibinfo{person}{K. Yang}, \bibinfo{person}{M. Choi},
  \bibinfo{person}{Z. Foo}, \bibinfo{person}{S. Bang}, \bibinfo{person}{S. Oh},
  \bibinfo{person}{D. Sylvester}, {and} \bibinfo{person}{D. Blaauw}.}
  \bibinfo{year}{2016}\natexlab{}.
\newblock \showarticletitle{8.5 A 60\%-efficiency 20nW-500µW tri-output fully
  integrated power management unit with environmental adaptation and
  load-proportional biasing for IoT systems}. In \bibinfo{booktitle}{\emph{2016
  IEEE International Solid-State Circuits Conference (ISSCC)}}.
  \bibinfo{pages}{154--155}.
\newblock
\urldef\tempurl%
\url{https://doi.org/10.1109/ISSCC.2016.7417953}
\showDOI{\tempurl}


\bibitem[\protect\citeauthoryear{Jung, Oh, Bang, Lee, Foo, Kim, Zhang,
  Sylvester, and Blaauw}{Jung et~al\mbox{.}}{2014}]%
        {hrv}
\bibfield{author}{\bibinfo{person}{W. Jung}, \bibinfo{person}{S. Oh},
  \bibinfo{person}{S. Bang}, \bibinfo{person}{Y. Lee}, \bibinfo{person}{Z.
  Foo}, \bibinfo{person}{G. Kim}, \bibinfo{person}{Y. Zhang},
  \bibinfo{person}{D. Sylvester}, {and} \bibinfo{person}{D. Blaauw}.}
  \bibinfo{year}{2014}\natexlab{}.
\newblock \showarticletitle{An Ultra-Low Power Fully Integrated Energy
  Harvester Based on Self-Oscillating Switched-Capacitor Voltage Doubler}.
\newblock \bibinfo{journal}{\emph{IEEE Journal of Solid-State Circuits}}
  \bibinfo{volume}{49}, \bibinfo{number}{12} (\bibinfo{year}{2014}),
  \bibinfo{pages}{2800--2811}.
\newblock
\urldef\tempurl%
\url{https://doi.org/10.1109/JSSC.2014.2346788}
\showDOI{\tempurl}


\bibitem[\protect\citeauthoryear{Kim, Lee, Foo, Pannuto, Kuo, Kempke, Ghaed,
  Bang, Lee, Kim, Jeong, Dutta, Sylvester, and Blaauw}{Kim
  et~al\mbox{.}}{2014}]%
        {mmimage}
\bibfield{author}{\bibinfo{person}{G. Kim}, \bibinfo{person}{Y. Lee},
  \bibinfo{person}{Zhiyoong Foo}, \bibinfo{person}{P. Pannuto},
  \bibinfo{person}{Ye-Sheng Kuo}, \bibinfo{person}{B. Kempke},
  \bibinfo{person}{M.~H. Ghaed}, \bibinfo{person}{Suyoung Bang},
  \bibinfo{person}{Inhee Lee}, \bibinfo{person}{Yejoong Kim},
  \bibinfo{person}{Seokhyeon Jeong}, \bibinfo{person}{P. Dutta},
  \bibinfo{person}{D. Sylvester}, {and} \bibinfo{person}{D. Blaauw}.}
  \bibinfo{year}{2014}\natexlab{}.
\newblock \showarticletitle{A millimeter-scale wireless imaging system with
  continuous motion detection and energy harvesting}. In
  \bibinfo{booktitle}{\emph{2014 Symposium on VLSI Circuits Digest of Technical
  Papers}}. \bibinfo{pages}{1--2}.
\newblock
\urldef\tempurl%
\url{https://doi.org/10.1109/VLSIC.2014.6858425}
\showDOI{\tempurl}


\bibitem[\protect\citeauthoryear{Lee, Mok, Huang, Cui, Lee, Leung, Mercier,
  Shellhammer, Larson, Asbeck, Rao, Song, Nurmikko, and Laiwalla}{Lee
  et~al\mbox{.}}{2019}]%
        {imd}
\bibfield{author}{\bibinfo{person}{J. Lee}, \bibinfo{person}{E. Mok},
  \bibinfo{person}{J. Huang}, \bibinfo{person}{L. Cui}, \bibinfo{person}{A.
  Lee}, \bibinfo{person}{V. Leung}, \bibinfo{person}{P. Mercier},
  \bibinfo{person}{S. Shellhammer}, \bibinfo{person}{L. Larson},
  \bibinfo{person}{P. Asbeck}, \bibinfo{person}{R. Rao}, \bibinfo{person}{Y.
  Song}, \bibinfo{person}{A. Nurmikko}, {and} \bibinfo{person}{F. Laiwalla}.}
  \bibinfo{year}{2019}\natexlab{}.
\newblock \showarticletitle{An Implantable Wireless Network of Distributed
  Microscale Sensors for Neural Applications}. In
  \bibinfo{booktitle}{\emph{2019 9th International IEEE/EMBS Conference on
  Neural Engineering (NER)}}. \bibinfo{pages}{871--874}.
\newblock
\urldef\tempurl%
\url{https://doi.org/10.1109/NER.2019.8717023}
\showDOI{\tempurl}


\bibitem[\protect\citeauthoryear{Lefebvre, Moreau, Dekimpe, and Bol}{Lefebvre
  et~al\mbox{.}}{2021}]%
        {imagersoc}
\bibfield{author}{\bibinfo{person}{M. Lefebvre}, \bibinfo{person}{L. Moreau},
  \bibinfo{person}{R. Dekimpe}, {and} \bibinfo{person}{D. Bol}.}
  \bibinfo{year}{2021}\natexlab{}.
\newblock \showarticletitle{7.7 A 0.2-to-3.6TOPS/W Programmable Convolutional
  Imager SoC with In-Sensor Current-Domain Ternary-Weighted MAC Operations for
  Feature Extraction and Region-of-Interest Detection}. In
  \bibinfo{booktitle}{\emph{2021 IEEE International Solid- State Circuits
  Conference (ISSCC)}}, Vol.~\bibinfo{volume}{64}. \bibinfo{pages}{118--120}.
\newblock
\urldef\tempurl%
\url{https://doi.org/10.1109/ISSCC42613.2021.9365839}
\showDOI{\tempurl}


\bibitem[\protect\citeauthoryear{Lin, Maire, Belongie, Hays, Perona, Ramanan,
  Dollár, and Zitnick}{Lin et~al\mbox{.}}{2014}]%
        {coco-dataset}
\bibfield{author}{\bibinfo{person}{T. Lin}, \bibinfo{person}{M Maire},
  \bibinfo{person}{S. Belongie}, \bibinfo{person}{J. Hays}, \bibinfo{person}{P.
  Perona}, \bibinfo{person}{D. Ramanan}, \bibinfo{person}{P. Dollár}, {and}
  \bibinfo{person}{C.~L. Zitnick}.} \bibinfo{year}{2014}\natexlab{}.
\newblock \showarticletitle{Microsoft COCO: Common Objects in Context}. In
  \bibinfo{booktitle}{\emph{European Conference on Computer Vision (ECCV)}}
  (2014-01-01). \bibinfo{address}{Zürich}.
\newblock
\urldef\tempurl%
\url{cocodataset.org}
\showURL{%
\tempurl}
\newblock
\shownote{Oral.}


\bibitem[\protect\citeauthoryear{Lu, Le, and Kim}{Lu et~al\mbox{.}}{2021}]%
        {realtimegr}
\bibfield{author}{\bibinfo{person}{Y. Lu}, \bibinfo{person}{V.~L. Le}, {and}
  \bibinfo{person}{T.~T. Kim}.} \bibinfo{year}{2021}\natexlab{}.
\newblock \showarticletitle{9.7 A 184 µ W Real-Time Hand-Gesture Recognition
  System with Hybrid Tiny Classifiers for Smart Wearable Devices}. In
  \bibinfo{booktitle}{\emph{2021 IEEE International Solid- State Circuits
  Conference (ISSCC)}}, Vol.~\bibinfo{volume}{64}. \bibinfo{pages}{156--158}.
\newblock
\urldef\tempurl%
\url{https://doi.org/10.1109/ISSCC42613.2021.9365963}
\showDOI{\tempurl}


\bibitem[\protect\citeauthoryear{Moon, Lee, Blaauw, and Phillips}{Moon
  et~al\mbox{.}}{2019}]%
        {sol}
\bibfield{author}{\bibinfo{person}{E. Moon}, \bibinfo{person}{I. Lee},
  \bibinfo{person}{D. Blaauw}, {and} \bibinfo{person}{J.D. Phillips}.}
  \bibinfo{year}{2019}\natexlab{}.
\newblock \showarticletitle{High-efficiency photovoltaic modules on a chip for
  millimeter-scale energy harvesting}.
\newblock \bibinfo{journal}{\emph{Progress in Photovoltaics: Research and
  Applications}} \bibinfo{volume}{27}, \bibinfo{number}{6}
  (\bibinfo{year}{2019}), \bibinfo{pages}{540--546}.
\newblock
\urldef\tempurl%
\url{https://doi.org/10.1002/pip.3132}
\showDOI{\tempurl}


\bibitem[\protect\citeauthoryear{Morishita, Kato, Okubo, Toi, Hiraki, Otani,
  Abe, Shinohara, and Kondo}{Morishita et~al\mbox{.}}{2021}]%
        {imagesensorandacc}
\bibfield{author}{\bibinfo{person}{F. Morishita}, \bibinfo{person}{N. Kato},
  \bibinfo{person}{S. Okubo}, \bibinfo{person}{T. Toi}, \bibinfo{person}{M.
  Hiraki}, \bibinfo{person}{S. Otani}, \bibinfo{person}{H. Abe},
  \bibinfo{person}{Y. Shinohara}, {and} \bibinfo{person}{H. Kondo}.}
  \bibinfo{year}{2021}\natexlab{}.
\newblock \showarticletitle{A CMOS Image Sensor and an AI Accelerator for
  Realizing Edge-Computing-Based Surveillance Camera Systems}. In
  \bibinfo{booktitle}{\emph{2021 Symposium on VLSI Circuits}}.
  \bibinfo{pages}{1--2}.
\newblock
\urldef\tempurl%
\url{https://doi.org/10.23919/VLSICircuits52068.2021.9492514}
\showDOI{\tempurl}


\bibitem[\protect\citeauthoryear{Naderiparizi, Zhang, Philipose, Priyantha,
  Liu, and Ganesan}{Naderiparizi et~al\mbox{.}}{2017}]%
        {continuouscv}
\bibfield{author}{\bibinfo{person}{S. Naderiparizi}, \bibinfo{person}{P.
  Zhang}, \bibinfo{person}{M. Philipose}, \bibinfo{person}{B. Priyantha},
  \bibinfo{person}{J. Liu}, {and} \bibinfo{person}{D. Ganesan}.}
  \bibinfo{year}{2017}\natexlab{}.
\newblock \showarticletitle{Glimpse: A Programmable Early-Discard Camera
  Architecture for Continuous Mobile Vision}. In
  \bibinfo{booktitle}{\emph{Proceedings of the 15th Annual International
  Conference on Mobile Systems, Applications, and Services}} (Niagara Falls,
  New York, USA) \emph{(\bibinfo{series}{MobiSys '17})}.
  \bibinfo{publisher}{Association for Computing Machinery},
  \bibinfo{address}{New York, NY, USA}, \bibinfo{pages}{292–305}.
\newblock
\showISBNx{9781450349284}
\urldef\tempurl%
\url{https://doi.org/10.1145/3081333.3081347}
\showDOI{\tempurl}


\bibitem[\protect\citeauthoryear{{Pannuto}, {Lee}, {Kuo}, {Foo}, {Kempke},
  {Kim}, {Dreslinski}, {Blaauw}, and {Dutta}}{{Pannuto} et~al\mbox{.}}{2015}]%
        {mbus}
\bibfield{author}{\bibinfo{person}{P. {Pannuto}}, \bibinfo{person}{Y. {Lee}},
  \bibinfo{person}{Y. {Kuo}}, \bibinfo{person}{Z. {Foo}}, \bibinfo{person}{B.
  {Kempke}}, \bibinfo{person}{G. {Kim}}, \bibinfo{person}{R.~G. {Dreslinski}},
  \bibinfo{person}{D. {Blaauw}}, {and} \bibinfo{person}{P. {Dutta}}.}
  \bibinfo{year}{2015}\natexlab{}.
\newblock \showarticletitle{MBus: An ultra-low power interconnect bus for next
  generation nanopower systems}. In \bibinfo{booktitle}{\emph{2015 ACM/IEEE
  42nd Annual International Symposium on Computer Architecture (ISCA)}}.
  \bibinfo{pages}{629--641}.
\newblock


\bibitem[\protect\citeauthoryear{Rossi, Conti, Eggiman, Mach, Mauro, Guermandi,
  Tagliavini, Pullini, Loi, Chen, Flamand, and Benini}{Rossi
  et~al\mbox{.}}{2021}]%
        {iotsoc}
\bibfield{author}{\bibinfo{person}{D. Rossi}, \bibinfo{person}{F. Conti},
  \bibinfo{person}{M. Eggiman}, \bibinfo{person}{S. Mach},
  \bibinfo{person}{A.~D. Mauro}, \bibinfo{person}{M. Guermandi},
  \bibinfo{person}{G. Tagliavini}, \bibinfo{person}{A. Pullini},
  \bibinfo{person}{I. Loi}, \bibinfo{person}{J. Chen}, \bibinfo{person}{E.
  Flamand}, {and} \bibinfo{person}{L. Benini}.}
  \bibinfo{year}{2021}\natexlab{}.
\newblock \showarticletitle{4.4 A 1.3TOPS/W @ 32GOPS Fully Integrated 10-Core
  SoC for IoT End-Nodes with 1.7$\upmu$W Cognitive Wake-Up From MRAM-Based
  State-Retentive Sleep Mode}. In \bibinfo{booktitle}{\emph{2021 IEEE
  International Solid- State Circuits Conference (ISSCC)}},
  Vol.~\bibinfo{volume}{64}. \bibinfo{pages}{60--62}.
\newblock
\urldef\tempurl%
\url{https://doi.org/10.1109/ISSCC42613.2021.9365939}
\showDOI{\tempurl}


\bibitem[\protect\citeauthoryear{Smets, Goedemé, Mittal, and Verhelst}{Smets
  et~al\mbox{.}}{2019}]%
        {realtimeimageproc}
\bibfield{author}{\bibinfo{person}{S. Smets}, \bibinfo{person}{T. Goedemé},
  \bibinfo{person}{A. Mittal}, {and} \bibinfo{person}{M. Verhelst}.}
  \bibinfo{year}{2019}\natexlab{}.
\newblock \showarticletitle{2.2 A 978GOPS/W Flexible Streaming Processor for
  Real-Time Image Processing Applications in 22nm FDSOI}. In
  \bibinfo{booktitle}{\emph{2019 IEEE International Solid- State Circuits
  Conference - (ISSCC)}}. \bibinfo{pages}{44--46}.
\newblock
\urldef\tempurl%
\url{https://doi.org/10.1109/ISSCC.2019.8662346}
\showDOI{\tempurl}


\bibitem[\protect\citeauthoryear{Song, Ding, Tiurin, Xu, Allebes, Singh, Zhang,
  Traferro, Korpela, Van~Helleputte, Staszewski, Liu, and Bachmann}{Song
  et~al\mbox{.}}{2020}]%
        {crystallesstx}
\bibfield{author}{\bibinfo{person}{M. Song}, \bibinfo{person}{M. Ding},
  \bibinfo{person}{E. Tiurin}, \bibinfo{person}{K. Xu}, \bibinfo{person}{E.
  Allebes}, \bibinfo{person}{G. Singh}, \bibinfo{person}{P. Zhang},
  \bibinfo{person}{S. Traferro}, \bibinfo{person}{H. Korpela},
  \bibinfo{person}{N. Van~Helleputte}, \bibinfo{person}{R.~B. Staszewski},
  \bibinfo{person}{Y. Liu}, {and} \bibinfo{person}{C. Bachmann}.}
  \bibinfo{year}{2020}\natexlab{}.
\newblock \showarticletitle{30.8 A 3.5mm×3.8mm Crystal-Less MICS Transceiver
  Featuring Coverages of ±160ppm Carrier Frequency Offset and 4.8-VSWR Antenna
  Impedance for Insertable Smart Pills}. In \bibinfo{booktitle}{\emph{2020 IEEE
  International Solid- State Circuits Conference - (ISSCC)}}.
  \bibinfo{pages}{474--476}.
\newblock
\urldef\tempurl%
\url{https://doi.org/10.1109/ISSCC19947.2020.9063083}
\showDOI{\tempurl}


\bibitem[\protect\citeauthoryear{Thijssen, Klumperink, Quinlan, and
  Nauta}{Thijssen et~al\mbox{.}}{2020}]%
        {ulpblerx}
\bibfield{author}{\bibinfo{person}{B.~J. Thijssen}, \bibinfo{person}{E.~A.~M.
  Klumperink}, \bibinfo{person}{P. Quinlan}, {and} \bibinfo{person}{B. Nauta}.}
  \bibinfo{year}{2020}\natexlab{}.
\newblock \showarticletitle{30.4 A 370µW 5.5dB-NF BLE/BT5.0/IEEE
  802.15.4-Compliant Receiver with gt;63dB Adjacent Channel Rejection at gt;2
  Channels Offset in 22nm FDSOI}. In \bibinfo{booktitle}{\emph{2020 IEEE
  International Solid- State Circuits Conference - (ISSCC)}}.
  \bibinfo{pages}{466--468}.
\newblock
\urldef\tempurl%
\url{https://doi.org/10.1109/ISSCC19947.2020.9062973}
\showDOI{\tempurl}


\bibitem[\protect\citeauthoryear{T.Toutin}{T.Toutin}{2004}]%
        {distortion-3}
\bibfield{author}{\bibinfo{person}{T.Toutin}.} \bibinfo{year}{2004}\natexlab{}.
\newblock \showarticletitle{Review article: Geometric processing of remote
  sensing images: models, algorithms and methods}.
\newblock \bibinfo{journal}{\emph{International Journal of Remote Sensing}}
  \bibinfo{volume}{25}, \bibinfo{number}{10} (\bibinfo{year}{2004}),
  \bibinfo{pages}{1893--1924}.
\newblock
\urldef\tempurl%
\url{https://doi.org/10.1080/0143116031000101611}
\showDOI{\tempurl}


\bibitem[\protect\citeauthoryear{Xu, Li, Lin, Wei, Qiao, Yin, and Yang}{Xu
  et~al\mbox{.}}{2021}]%
        {senseandcompute}
\bibfield{author}{\bibinfo{person}{H. Xu}, \bibinfo{person}{Z. Li},
  \bibinfo{person}{N. Lin}, \bibinfo{person}{Q. Wei}, \bibinfo{person}{F.
  Qiao}, \bibinfo{person}{X. Yin}, {and} \bibinfo{person}{H. Yang}.}
  \bibinfo{year}{2021}\natexlab{}.
\newblock \showarticletitle{MACSen: A Processing-In-Sensor Architecture
  Integrating MAC Operations Into Image Sensor for Ultra-Low-Power BNN-Based
  Intelligent Visual Perception}.
\newblock \bibinfo{journal}{\emph{IEEE Transactions on Circuits and Systems II:
  Express Briefs}} \bibinfo{volume}{68}, \bibinfo{number}{2}
  (\bibinfo{year}{2021}), \bibinfo{pages}{627--631}.
\newblock
\urldef\tempurl%
\url{https://doi.org/10.1109/TCSII.2020.3015902}
\showDOI{\tempurl}


\end{thebibliography}
\end{document}